\documentstyle{l-aa}
\input{epsf}
\begin{document}

\title{ Neutrino pair emission due to scattering of electrons
        off fluxoids in superfluid neutron star cores}
\author{ A.D.~Kaminker$^1$, D.G.~Yakovlev$^1$,
              and P.~Haensel$^2$}
\offprints{P.~Haensel}
\institute{A.F.~Ioffe Institute of Physics and Technology,
              194021 St.Petersburg, Russia
              \and
              N.~Copernicus Astronomical Center,
              Polish Academy of Sciences, Bartycka 18,
              00-716 Warszawa, Poland}
\thesaurus{02.04.1 - 08.09.3 - 08.14.1}
\maketitle
\markboth{A.D. Kaminker et al: Neutrino pair emission}{}
\label{sampout}

\def\la{\;
\raise0.3ex\hbox{$<$\kern-0.75em\raise-1.1ex\hbox{$\sim$}}\; }
\def\ga{\;
\raise0.3ex\hbox{$>$\kern-0.75em\raise-1.1ex\hbox{$\sim$}}\; }

\begin{abstract}
We study the emission of neutrinos, resulting
from the scattering of electrons off magnetic flux tubes (fluxoids)
in the neutron star cores with superfluid
(superconducting) protons. In the absence of proton superfluidity
($T \geq T_{cp}$), this process transforms into
the well known electron synchrotron emission of neutrino pairs
in a locally uniform magnetic field $B$, with the neutrino energy
loss rate $Q$ proportional to $B^2 T^5$. For
temperatures $T$ not  much below $T_{cp}$, the synchrotron
regime ($Q \propto T^5$) persists and the emissivity
$Q$ can be amplified by several orders of magnitude due to
the appearance of the fluxoids
and associated enhancement of the field within them.
For lower $T$, the synchrotron regime transforms into
the bremsstrahlung regime ($Q \propto T^6$) similar
to the ordinary neutrino-pair bremsstrahlung of electrons
which scatter off atomic nuclei.
We calculate $Q$ numerically and
represent our results through a suitable  analytic fit.
In addition, we estimate the emissivities of
two other neutrino-production mechanisms
which are usually neglected -- neutrino-pair
bremsstrahlung processes due to electron-proton and electron-electron
collisions. We show that the electron-fluxoid
and electron-electron scattering
can provide the main neutrino production mechanisms
in the neutron star cores
with highly superfluid protons and neutrons
at $T \la 5 \times 10^8$ K. The electron-fluxoid scattering
is significant if the initial, locally uniform magnetic field
$B \ga 10^{13}$ G.
\end{abstract}

\section{Introduction}
Neutrino energy losses play a decisive role in cooling of young
neutron stars (NSs).
During first $10^5-10^6~$ yrs after the NS birth,
the cooling is dominated by the
neutrino emission from the NS cores, of
density exceeding the normal nuclear matter density,
$\rho_0=2.7 \times 10^{14}$ g~cm$^{-3}$ (corresponding to the
nucleon number density  $n_0=0.16~{\rm fm^{-3}}$).

The emission of neutrinos results
from the weak interaction processes involving baryons and/or
leptons in the NS cores. The presence of the
strong internal magnetic field allows an
additional process --- the neutrino synchrotron
radiation of electrons. This process was considered by many authors
(e.g, Kaminker et al. 1991, 1992, and references therein) for
a spatially homogeneous magnetic field.

It is widely accepted (see, e.g.,
Takatsuka \& Tamagaki 1993, and references therein), that
neutrons and/or protons in the NS core can be superfluid.
Let $T_{cp}$ and $T_{cn}$ be the critical temperatures
of the proton and neutron superfluidities, respectively.
Theoretical values of $T_{cp}$ and $T_{cn}$ are strongly  model dependent
and range from $10^8$ to $10^{10}$~K.
For $T \ll T_c$, the superfluidity of nucleons
 strongly suppresses the neutrino luminosity produced by the
Urca and bremsstrahlung processes in the
NS cores (e.g., Pethick 1992, Yakovlev \& Levenfish 1995).
This enhances the relative importance of the neutrino
synchrotron process.

It is commonly thought that superfluid protons form
the type II superconductor (see Sauls 1989, and references therein).
Let us assume that the core of a newly born,
hot and initially nonsuperfluid NS
contains a magnetic field (e.g., primordial one,
or generated at the NS formation stage).
The transition to a superconducting state
in the course of the stellar cooling would then be
accompanied by a dramatic change in the spatial structure of the
magnetic field. Initially homogeneous field
splits into an ensemble
of fluxoids which contain
a superstrong magnetic field, embedded in the field-free
superconducting medium. So far, the
neutrino synchrotron radiation has been studied
 only for a homogeneous magnetic field. In the
present paper, we show that
after the superfluidity onset,  the neutrino synchrotron mechanism
transforms
into the neutrino pair emission   due to the electron-fluxoid
scattering.

Physical conditions in the superconducting NS cores are
described in Sect.~2. In Sect.~3, we present general
formalism of the
electron-fluxoid ($ef$) scattering. The calculations
of the neutrino emissivity
are performed in
Sect.~4. In Sect.~5, we estimate
the emissivities of two additional
neutrino production mechanisms which are usually neglected:
the bremsstrahlung processes due to electron-proton ($ep$) and
electron-electron ($ee$) scattering.
No calculation of
these emissivities has been done earlier, to our knowledge, although
the importance of the $ep$ scattering has been
mentioned by Schaab et al. (1996).
In Sect.~6 we show that the new mechanisms (especially $ef$
and $ee$ scattering)
can be important in the NS cores
with highly superfluid protons and neutrons since the superfluidity
reduces strongly the familiar neutrino generation
reactions (Urca processes, nucleon-nucleon bremsstrahlung).
%
\section{Superconducting neutron star cores}
The density within a NS
core is expected to range from about 0.5$\rho_0$
at its outer edge (Lorenz et al. 1993),
to (5--10)$\rho_0$ at the star center.
For densities $\rho \la 2 \rho_0$, matter
consists of neutrons, with a few percent admixture of protons,
electrons, and possibly muons.
The fraction of muons is usually much smaller
than that of electrons, and a simple $npe$  model
is a good approximation. At higher densities,
$\rho \ga 2 \rho_0$, other particles may appear, such as
hyperons, condensed pions and/or kaons, or even free quarks.
Dependence of the neutrino emissivity on
the composition of dense matter is
reviewed by Pethick (1992).
For simplicity, we will neglect
exotic matter constituents, restricting
ourselves to the simplest $npe$ model.
All constituents of the $npe$  matter are degenerate. The electrons
 form an ultrarelativistic and almost ideal gas, with
the electron Fermi energy $\mu_e \approx \hbar c (3 \pi^2 n_e)^{1/3} \sim$
(100--300) MeV,
where $n_e$ is the electron number density.
The neutrons and protons form strongly interacting Fermi
liquids. The Fermi energy of neutrons is of the order
of that of the electrons, while the Fermi energy of protons
is much smaller.

Both nucleon species, $n$ and $p$, can be in a superfluid
state (for a review, see Takatsuka \& Tamagaki 1993).
The superfluidity is widely accepted to be of
the BCS--type; the $nn$ and/or $pp$ pairing occurs due
to nuclear attraction. Let us focus on the
proton superfluidity.
In view of a relatively low proton number density,
the inter-proton distance
is rather high, and the $pp$ interaction is attractive
in the $^1S_0$ state. Thus, the proton pairing is most
likely to be in the $^1S_0$ state. Recent reviews on
the properties of superconducting protons in the
NS cores are given by
Sauls (1989) and  Bhattacharya \& Srinivasan  (1995). The
 proton superconductivity is characterized by an
isotropic energy gap in the single-(quasi)particle spectrum
$\Delta_p$, which depends on temperature and density.
The critical temperature $T_{cp}$
is related to the zero-temperature energy gap $\Delta_p(0)$
by $T_{cp}=\Delta_p(0)/(1.76 \, k_{\rm B})$
(Lifshitz \& Pitaevskii 1980), where
$k_{\rm B}$ is the Boltzmann constant.
An important parameter of
superconducting protons is the $pp$ correlation length
$\xi$; it measures the size of a $pp$ Cooper pair.
In the BCS model, $\xi$ is related to $\Delta_p$
and the proton Fermi velocity $v_{{\rm F}p}$ by
$\xi= \hbar v_{{\rm F}p}/ ( \pi \Delta_p)$.
The zero temperature value of $\xi$ will be denoted by $\xi_0$.
Another important parameter is the penetration depth $\lambda$
of the magnetic field into a proton superconductor.
In the case of $\lambda \gg \xi$ (which corresponds to the
conditions prevailing in the NS cores), the zero
temperature penetration depth, $\lambda_0$, is determined by the
proton plasma frequency, $\omega_p$
(e.g., Lifshitz \& Pitaevskii 1980),
\begin{equation}
\lambda_0={c\over\omega_p}~,
\hspace{10mm}
\omega_p =
\left({4\pi n_p e^2\over  m_p^\ast}\right)^{1/2}~,
\label{lambda_0}
\end{equation}
where $n_p$ is the proton number density, and
$m_p^\ast$ is the proton effective mass,
that can differ from the bare mass $m_p$
due to the many-body effects in dense matter.
Simple estimates yield typical values of
$\xi_0$ of a few fm, and $\lambda_0$ of a few ten fm.
If so, $\lambda_0 \gg \xi_0$,  which means that protons
constitute a type II superconductor. Notice that the values of
$\lambda_0$ and $\xi_0$ are model dependent, and one cannot
exclude the case of $\lambda_0 \sim \xi_0$, in which the
proton superconductivity could be of type I, although we will not
consider this case in the present article (some comments on the 
case of a type I proton superconductor  are given at the end 
of Sect. 6).

For $T \to T_{cp}$, both $\xi$ and $\lambda$ diverge
as $(T_{cp}-T)^{-{1/2}}$, while for $T\ll T_{cp}$ they can be
replaced by their zero temperature values, $\xi_0$ and
$\lambda_0$. However,  we need to
know the temperature dependence of $\lambda$ for all
temperatures below $T_{cp}$. In what follows, we will
approximate
   its
temperature dependence
   by the
Gorter--Casimir formula (Tilley \& Tilley 1990),
\begin{equation}
\lambda={ \lambda_0 \over \sqrt{1 - (T/T_{cp})^4}}~.
\label{lambda}
\end{equation}
%

The transition to the type II superconductivity
during the NS cooling
is accompanied by the formation of
quantized flux tubes (Abrikosov fluxoids), parallel to
the initial local magnetic field $\bar{\bf B}$.
Each fluxoid carries an
elementary magnetic flux $\phi_0= \pi \hbar c/e$.
The number of fluxoids per unit area perpendicular
to the initial field is
${\cal N}_{\rm F}= \bar{B} / \phi_0$,
and the mean distance
between the fluxoids is
$d_{\rm F} = [2\phi_0 /( \sqrt{3} \bar{B})]^{1/2}$.
Simple estimate yields
$d_{\rm F} \approx 1500 / (\bar{B}_{13})^{1/2}~{\rm fm}$,
where $\bar{B}_{13} \equiv \bar{B} /(10^{13}~{\rm G})$. A fluxoid
has a small central core
of radius $\sim \xi$ containing normal protons.
  A typical fluxoid radius is
$\lambda$. Just after the
superconductivity onset
  ($T < T_{cp}$), this radius
is large and the fluxoids
fill all the available space. When temperature
  drops
to about $0.8 T_{cp}$, $\lambda$  reduces nearly
to its zero-temperature value $\lambda_0$. Thus, $\lambda$
becomes much
smaller than the inter-fluxoid distance $d_{\rm F}$, for
the magnetic fields $\bar{B} < 10^{15}$ G which we
will consider in the present article.
The maximum value of $B$ is reached at the
fluxoid axis, $B_{\rm max} \simeq$
$[\phi_0 /( \pi \lambda^2)]\, \log(\lambda/\xi)$.
In our case $\lambda \gg \xi$, and the
magnetic field
   profile at  $r\gg \xi$
is given by
(e.g., Lifshitz \& Pitaevskii 1980):
\begin{equation}
           B(r)={\phi_0\over 2\pi\lambda^2}
           K_0\left({r\over \lambda}\right),
\label{Br}
\end{equation}
where
   $K_0(x)$
is a McDonald function.
In particular, for $r \gg \lambda$,
   one has
$B(r) \approx $
$\phi_0 (8\pi r\lambda^3)^{-1/2} \, \exp(-r/\lambda)$.

The superconducting state is destroyed when
$d_{\rm F} < \xi$, which corresponds to magnetic fields $\bar{B}>
B_{c2}=\phi_0/(\pi\xi^2)$, whose typical values are
$\sim 10^{18}$ G.
We do not consider such  strong fields.

Let us mention that
the fluxoids may migrate slowly
   outward the stellar core
due to the buoyancy forces
(Muslimov \& Tsygan 1985, Jones 1987, Srinivasan
et al.\ 1990).

\section{General formalism}
Consider the neutrino-pair emission
due to scattering of strongly degenerate, relativistic
electrons off fluxoids (Sect.~2),
\begin{equation}
         e + f \to e + f + \nu + \bar{\nu}.
\label{ef}
\end{equation}
We will treat this process using the standard
perturbation theory with free electrons in nonperturbed states.
The process (\ref{ef}) is similar to
the well known neutrino-pair bremsstrahlung
due to scattering of electrons by atomic nuclei (see, e.g.,
Festa and Ruderman 1969, Soyeur and Brown 1979,
Haensel et al. 1996, and references therein).
It is described by two second-order diagrams,
where one (electromagnetic) vertex is associated
with electron scattering by the fluxoid magnetic fields,
while the other (four-tail) vertex is due to
the neutrino-pair emission.

In what follows, we will mainly use the units in which
$c=\hbar=k_{\rm B}=1$ although we will insert the ordinary
physical units in the final expressions.

The neutrino energy loss rate (emissivity)  $Q_{\rm flux}$
  (ergs cm~$^{-3}$ s~$^{-1}$) of process (\ref{ef})
can be written as
\begin{eqnarray}
      Q_{\rm flux} & = & {{\cal N}_{\rm F} \over (2 \pi)^{10}}
            \int \! {\rm d} {\bf p} \! \int \! {\rm d} {\bf p}'
            \! \int \! {\rm d} {\bf k}_{\nu} \!
            \int \! {\rm d} {\bf k}'_{\nu}
            \nonumber \\
            & \times &
            \delta(\varepsilon - \varepsilon' -\omega) \,
            \delta(p_z - p'_z - k_z) \,
            \omega f(1-f') W ,
\label{Qgeneral}
\end{eqnarray}
where ${\cal N}_{\rm F}$
 is the fluxoid surface number density defined in Sect.\ 2,
$P = (\varepsilon, {\bf p})$ and
$P'= (\varepsilon',{\bf p}')$ are, respectively,
the electron 4-momenta in
the initial and final states,
$K_{\nu} = (\omega_{\nu}, {\bf k}_{\nu})$  and
$K'_{\nu} =(\omega'_{\nu}, {\bf k}'_{\nu})$
are the 4-momenta of the neutrino and of the
antineutrino, and
$K = K_{\nu} + K'_{\nu} = (\omega, {\bf k})$ is the 4-momentum
of the neutrino pair $(\omega = \omega_{\nu} + \omega'_{\nu}$ and
${\bf k} = {\bf k}_{\nu} + {\bf k}'_{\nu})$.
Furthermore,
\begin{equation}
       f = \left[ 1 + \exp \left( \frac{\varepsilon - \mu}{T}
                                \right) \right]^{-1}
\label{FDf}
\end{equation}
is the Fermi-Dirac function for the initial electron state, and
$f' \equiv f(\varepsilon')$ is the same function
for the final electron state.
The $\delta$-functions describe
energy conservation and momentum
conservation along the fluxoid axis
  (the axis $z$). Finally,
$W$ is the differential transition rate
\begin{equation}
      W= {G_{\rm F}^2 \over 2} \; \frac{1}
         {(2 \omega_\nu)(2 \omega'_\nu)
         (2 \varepsilon)(2 \varepsilon')} \;
         \sum_{\rm \sigma,\nu} | M |^2,
\label{W}
\end{equation}
where $G_{\rm F}= 1.436 \times 10^{-49}$ erg cm$^3$
  is the Fermi weak interaction constant,
and
$|M|^2$ is the squared transition matrix element. Summation is over the
electron spin states $\sigma$ before and after scattering
and over the neutrino flavors ($\nu_e$, $\nu_\mu$, $\nu_\tau$).
The neutrino energies are assumed to be much lower
than the intermediate boson mass
($\sim$80 GeV). Then the standard approach yields
\begin{eqnarray}
     \sum_{\sigma} |M|^2 & = & e^2 \,
           \int {\rm d} {\bf q} \,
          \delta({\bf p} - {\bf q} - {\bf p}' - {\bf k})
            A^i A^{j\ast} \,
\nonumber \\
     & \times &  {\rm Tr}(\hat{K}_{\nu} O^{\alpha} \hat{K}'_{\nu}
              O^{\beta})
              \nonumber  \\
              & \times &
              {\rm Tr} \left[ \bar{L}_{\beta j} (\hat{P}'+m_e) L_{\alpha i}
              (\hat{P}+m_e)  \right] ,
\label{M2} \\
    O^\alpha & = & \gamma^{\alpha} (1+ \gamma^5),
                 \nonumber  \\
    L_{\alpha i} & = & \Gamma_{\alpha} G(P-Q) \gamma_i
              +   \gamma_i G(P'+Q) \Gamma_{\alpha},
\label{OL} \\
    G(P) & = & \frac{\hat{P} + m_e}{P^2 - m_e^2},
              \hspace{3mm} \Gamma^{\alpha} = C_V \gamma^{\alpha} +
              C_A \gamma^{\alpha} \gamma^5.
\label{GG}
\end{eqnarray}
Here, ${\bf q} = {\bf p} - {\bf p}' - {\bf k}$ is a
momentum transfer (in the $(x,y)$-plane)
from an electron to a fluxoid,
$Q  = P - P' -K = (\Omega, {\bf q})$
is an appropriate 4-momentum transfer
(with no energy transfer, $\Omega=0$), ${\bf A} \equiv {\bf A}({\bf q})$ is
a 2-dimensional Fourier transform of the
magnetic-field vector-potential, which lies in the
$(x,y)$ plane, and $m_e$ is the electron mass.
Greek indices $\alpha$ and $\beta$ run over (0,1,2,3) and
Latin ones $i$ and $j$ refer to the spatial components (1,2,3);
$G(P)$ is the free-electron propagator,
$\gamma^\alpha$ is a Dirac
matrix, upper bar denotes Dirac conjugate, and
$\hat{P} \equiv P_\alpha \gamma^\alpha$
(Berestetskii et al. 1982).
Furthermore, $C_V$ and $C_A$ are
the vector and the axial vector
weak interaction constants, respectively. For the
emission of the electron neutrinos
(charged + neutral currents), one has
$C_V = 2 \sin^2 \theta_{\rm W} +0.5$ and $C_A= 0.5$,
while for the emission of the muonic or the tauonic
neutrinos (neutral currents only),
$C'_V = 2 \sin^2 \theta_{\rm W} - 0.5$ and $C'_A = -0.5$;
$\theta_{\rm W}$ is the Weinberg angle,
$\sin^2 \theta_{\rm W} \simeq 0.23$.

Using the identity (Berestetskii et al. 1982)
\begin{eqnarray}
           &  \,  &
           \int \! {\rm d} {\bf k}_\nu \int \! {\rm d} {\bf k}'_\nu \,
           \delta^{(4)}(K - K_\nu - K'_\nu) \,
           \frac{ K_\nu^{\alpha} {K'}_\nu^{\beta}}{\omega_\nu \omega'_\nu}
           \nonumber  \\
           &  \, &  \, = \, \frac{\pi}{ 6} (K^2 g^{\alpha \beta} +
           2 K^\alpha K^\beta),
\label{Identity}
\end{eqnarray}
we obtain
\begin{eqnarray}
        Q_{\rm flux} & = &
             \frac{e^2 \, G_{\rm F}^2 \,{\cal N}_{\rm F} }{12 (2 \pi)^{9}}
             \int  {\rm d} {\bf p} \int  {\rm d} {\bf p}'
             \int  {\rm d} {\bf k} \,
             \delta(p_z - p'_z - k_z)
\nonumber   \\
              & \times &  A^i A^{j\ast}
               \frac{\omega}{ \varepsilon \varepsilon'} \,
                J_{i j} \, f (1 - f') ,
\label{Q1} \\
          J _{i j}  & = & \sum_\nu (K^{\alpha} K^{\beta} -
              K^2 g^{\alpha \beta})
              \nonumber   \\
              & \times &
              {\rm Tr} \left[(\hat{P}' + m_e) L_{\alpha  i}
              (\hat{P} + m_e)
              \bar{L}_{\beta  j}  \right],
\label{Jgeneral}
\end{eqnarray}
where $g^{\alpha \beta}= {\rm diag}(1,-1,-1,-1)$ is the metric tensor.
The integration in Eq.\,(\ref{Q1}) is to be
carried out over the domain
    in which
$K^2 \geq 0$.
Notice that Eq.\,(\ref{Q1}) reproduces the well-known
expression for the neutrino-pair bremsstrahlung of electrons
which scatter off atomic nuclei (e.g., Haensel et al. 1996).
For this purpose, one needs only to replace
$
2\pi {\cal N}_{\rm F}e^2 \delta(p_z - p'_z - k_z)
A^i A^{j\ast} \to
n_{\rm i} |U({\bf q})|^2 \delta_{i j} \delta_{i 0},
$
where $n_{\rm i}$ is the number
density of nuclei and $U({\bf q})$ is the Fourier transform of the
electron-nucleus Coulomb potential.

The 2-dimensional Fourier
      transform $B(q)=B_z(q)$
of the
fluxoid magnetic field (\ref{Br})
in cylindrical coordinates is
\begin{eqnarray}
       B(q)  & = &
       \frac{\Phi_0}{\lambda^2}
       \int_0^{\infty} {\rm d} r \, r K_0
       \left( \frac{r}{\lambda}  \right)
        J_0 (qr) \, \, \,
        \frac{\Phi_0 q_0^2}{q^2 + q_0^2},
\label{Bq}
\end{eqnarray}
where $J_0(x)$ is a Bessel function and $q_0 =1/\lambda$.
Then, using cylindrical gauge, we have
${\bf A}({\bf q})  =$
$- i {\bf e}_A \, B(q)/q$,
  ${\bf e}_A = ({\bf B \times q})/(Bq)$ being
a unit vector.
Accordingly, in Eq.\,(\ref{Q1})
$A^i A^{j\ast} =$
$(B(q)/q)^2 \, e_A^i e_A^j $,
where $i,j$ = 1 or 2, for nonvanishing components.

Equations (\ref{Q1}) and (\ref{Jgeneral}) determine
the neutrino emissivity for any degree
of electron degeneracy and relativism.
We are interested in the case
of ultrarelativistic, strongly degenerate electrons (Sect.~2).
In the relativistic limit, tedious but straightforward
calculations yield:
\begin{equation}
          e_A^i e_A^j \, J_{i j} \approx  C_+^2 J_+ ,
        \, \, \, \,  C_+^2    =   \sum_\nu (C_V^2+ C_A^2),
\label{eAeAJ}
\end{equation}
\begin{eqnarray}
    J_+  & = & \frac{4 K^2}{uv} q^2 \left[ \, 2 ({\bf e}_A {\bf p})
               ({\bf e}_A {\bf p}')
               + \varepsilon \varepsilon'-{\bf p}{\bf p}' \, \right]
                \, +  \, 8K^2  \,
\nonumber    \\
          & - & \frac{4 K^2}{uv}
                 (\varepsilon \varepsilon' - {\bf p}{\bf p}')
                \left[ K^2 - 2 ({\bf e}_A {\bf k})^2  \right] \,
\nonumber     \\
             & + &  4 K^2 \frac{({\bf q} {\bf k})^2}{uv}\,
                + 4\, K^4 \frac{({\bf e}_A {\bf k})^2}{uv}
\nonumber     \\
          & - & 2 K^4 \left[  \,
                      2({\bf e}_A {\bf p})({\bf e}_A {\bf p}') +
                \varepsilon \varepsilon' - {\bf p} {\bf p}' \, \right]
                \left( \frac{1}{u} - \frac{1}{v} \right)^2
\nonumber      \\
           &  - & \frac{8 K^2}{uv} ({\bf q}
                  {\bf k})({\bf e}_A {\bf k})^2
\nonumber       \\
               &  +  &
              K^4 \left( \frac{1}{u^2} + \frac{1}{v^2} \right)
                (2{\bf q}{\bf k} - K^2) \,
\nonumber   \\
           &  -  &  4 K^4  ({\bf e}_A {\bf k})
                \left( \frac{{\bf e}_A {\bf p}}{u^2}
                - \frac{{\bf e}_A {\bf p}'}{v^2} \right),
\label{GenJ+}
\end{eqnarray}
where we
   introduce
the notations
\begin{eqnarray}
                u & =  &  \frac{1}{2} \left( P' + K \right)^2  =
                         P'K + \frac{1}{2} K^2 \, ,
\nonumber    \\
                v & = &   - \frac{1}{2} \left( P - K \right)^2  =
                         PK - \frac{1}{2} K^2 \, .
\label{uv}
\end{eqnarray}
Using  Eqs.\ (\ref{Q1}), (\ref{eAeAJ}),
and (\ref{GenJ+}), we get
\begin{eqnarray}
     Q_{\rm flux} & = & \frac{G_{\rm F}^2 e^2 C_+^2}
             {12 \, (2\pi)^9} \, {\cal N}_{\rm F}
              \int \, {\rm d} {\bf p} \int \, {\rm d} {\bf p}'
             \int \, {\rm d} {\bf k} \, \delta(p_z - p'_z - k_z)
\nonumber   \\
             & \times &
             \left( \frac{B(q)}{q} \right)^2 \;
             \frac{\omega }{\varepsilon \varepsilon'} \; J_+ \;
             f (1 - f')~ .
\label{Q2}
\end{eqnarray}

For practical applications, it is convenient to express
$Q_{\rm flux}$ in the form
\begin{eqnarray}
    Q_{\rm flux}  &  =  & \frac{ G_{\rm F}^2 e^2 \phi_0^2 C_+^2}
                          {2268 \, \hbar^9 c^8} \,
                          (k_{\rm B} T)^6 q_0
                          {\cal N}_{\rm F} \, L
\nonumber       \\
             & \, &  \,
\nonumber       \\
             & \approx  & 2.66  \times 10^{16} \, \bar{B}_{13} \, T_9^6 \,
                  \left( { n_p \, m_p \over n_0  \,
                   m_p^\ast }\right)^{1/2}
\nonumber      \\
              & \times &
             \left[ 1 - \left( {T  \over T_{cp}} \right)^4 \right]^{1/2}
             L \; \;  \; {\rm erg~s^{-1}~cm^{-3}}~.
\label{QthrL}
\end{eqnarray}
Here, $n_0 = 0.16$ fm$^{-3}$ is the standard
(nuclear)  number density,
$T_9$ is temperature in the units of $10^9$~K, and
$\bar{B}_{13}$ is defined
   in Sect.\ 2.
Numerical expression for
$Q_{\rm flux}$ takes into account
emission of $\nu_e$, $\nu_{\mu}$, $\nu_{\tau}$, so that
$C_+^2 \approx 1.675$. In Eq.\,(\ref{QthrL}),
we introduced the dimensionless quantity $L$, defined by
\begin{eqnarray}
     L & = & \frac{189}{(2\pi)^9 T^6 q_0 } \int \, {\rm d} {\bf p}  \,
         {\rm d} {\bf p}' \, {\rm d} {\bf k} \,
         \delta(p_z - p'_z - k_z)
\nonumber     \\
              &  \times &
               \left( {B(q) \over q \phi_0} \right)^2  \,
                   \frac{\omega}{\varepsilon \varepsilon'} \,
                    J_+ \, f (1 - f'),
\label{Lgeneral}
\end{eqnarray}
with $J_+$ given  by Eq.\,(\ref{GenJ+}).
Notice that Eq.\,(\ref{Lgeneral})
is valid for any axially symmetric
distribution of the fluxoid magnetic field $B(r)$ although we
will use a specific distribution, Eqs.\,(\ref{Br}) and (\ref{Bq}).
An analogous  quantity $L$
 was introduced in the case of the  neutrino-pair bremsstrahlung
due to the electron-nucleus scattering. In that case, it had a
 meaning of the
Coulomb logarithm (Haensel et al. 1996).

%
\section{Practical evaluation of $L$}
Let us calculate $L$ from Eq.\,(\ref{Lgeneral}).
Since the electrons are strongly
degenerate
  (Sect.\ 2),
the main contribution
 to $L$ comes from those electron transitions, in which
the electron momenta ${\bf p}$ and ${\bf p}'$ lie in the narrow
thermal shell around the Fermi surface, $|\varepsilon - \mu| \la T$
and $|\varepsilon' - \mu| \la T$. In Eq.\,(\ref{Lgeneral})
we may set
$
{\rm d} {\bf p} =
\varepsilon^2 \, {\rm d} \varepsilon \, {\rm d} \Omega
$,
where d$\Omega$ is a solid angle element in the direction
of {\bf p}. We also put $\varepsilon = p$ since the electrons
are ultrarelativistic
  (Sect.\ 2).
Then we can set
$\varepsilon = p_{\rm Fe}$ in all smooth functions of $\varepsilon$.
Afterwards, the integration over $\varepsilon$ is standard:
\begin{equation}
           \int \; {\rm d} \varepsilon \; f (1 - f') = \frac{\omega}
          {{\rm e}^{\omega/T} - 1}.
\label{Intdeps}
\end{equation}
The integrand in Eq.\,(\ref{Lgeneral})
depends obviously on the relative
azimuthal positions of {\bf p}, ${\bf p}'$ and {\bf k}.
Thus we can place {\bf p} in the
$(xz)$-plane without any loss of generality.

We shall mainly consider the case
of $q_0 \ll p_{\rm Fe}$ typical
for the NS cores (Sect.~2).
It is convenient to replace
the integration over ${\bf p}'$ in Eq.\,(\ref{Lgeneral})
by the integration over ${\bf q}$.
The integration is simplified because the
main contribution comes from the values of $q \ll p_{\rm Fe}$.
Thus we can use the approximation
of small-angle scattering.
Since $k$ can be comparable to $q$,
we should keep two first terms in Eq.\,(\ref{GenJ+}).
Using the inequalities $k \ll p$ and $q \ll p$ we obtain
\begin{equation}
              J_+  \approx 8 (\omega^2 - {\bf k}^2 )
              \left[ \frac{q^2}{q_r^2} \,
              \frac{({\bf e}_A {\bf p})^2}{p^2}  + \, 1 \right],
\label{J+sm}
\end{equation}
where we have neglected the quantity
$\varepsilon \varepsilon'- {\bf p}{\bf p}' =
[({\bf q} + {\bf k})^2 - \omega^2]/2$
in the first term of (\ref{GenJ+}), and have used
the approximate expressions $u \approx v \approx p q_r$,
which follow from Eqs.\ (\ref{uv}).
The subscript $r$ denotes the vector
component along ${\bf p}$; $q_r = q_x \, \sin \theta$,
where $\theta$ is an angle between {\bf p} and the $z$-axis
(electron pitch-angle).

Using energy---momentum conservation,
in analogy with the results of Haensel et al. (1996) we obtain
\begin{eqnarray}
            \omega  & = &  \varepsilon - \varepsilon'
            \approx  q_r + k_r,
\nonumber           \\
            K^2  & = &  \omega ^2 - k_r^2 - k_t^2   \approx
            k_0^2 - k_t^2,
\nonumber      \\
                  k_0^2   &  = &   q_r (2\omega -q_r),
\nonumber             \\
            J_+ & = & 8 (\omega^2 - {\bf k}^2) \frac{q^2}{q_r^2} \,
                   \sin^2 \theta ,
\label{omkJ+}
\end{eqnarray}
where $k_t$ is a component of {\bf k} transverse to {\bf p}.

The condition $\omega^2 \geq {\bf k}^2$ requires  $q_r >0$,
$\omega > q_r/2$ and $k_t^2 <  k_0^2 = q_r(2 \omega - q_r)$.
Taking into account
    Eq.\ (\ref{Intdeps}),
we obtain
\begin{eqnarray}
             L & = & { 189  \over  2^4 \pi^7 \, T^6 \, q_0 }
              \int_0^{\pi} {\rm d} \theta  \sin^3 \theta
              \int_0^\infty {\rm d} q_x
              \int_{-\infty}^\infty {\rm d} q_y
              \left( { B(q) \over \phi_0 \, q_r}
              \right)^2
\nonumber     \\
              & \, &  \,
\nonumber     \\
              & \times &
               \int_{q_r/2}^{\infty} {\rm d} \omega \,
              \frac{\omega^2}{{\rm e}^{\omega/T} -1 } \,
              \int_0^{k_0}  {\rm d} k_t \, k_t \, (k_0^2 - k_t^2).
\label{GenL}
\end{eqnarray}
The integrals converge rapidly
due to sharp decrease of $B(q)$, Eq.\,(\ref{Bq}).
In view of
  this,
we could extend the
integration over $q_x$ and $q_y$
up to $\infty$, and the lower integration limit
over $q_y$
  down
to $-\infty$.

Integrating over $k_t$, we have
\begin{eqnarray}
               L & = & {189 \over 8 \pi^7 \, T^6 \, q_0}
               \int_0^\pi {\rm d} \theta \,
               \sin^3  \theta  \int_0^\infty {\rm d} q_x
               \int_0^\infty
               {\rm d} q_y \,
\nonumber       \\
              &  \times  &
              \left( { B(q)  \over \phi_0} \right)^2
            \,  \int_{q_r/2}^{\infty} {\rm d} \omega \,
               \left( \omega - {q_r \over 2} \right)^2
               \, { \omega^2 \over {\rm e}^{\omega / T}-1},
\label{LRT}
\end{eqnarray}
which can be transformed as
\begin{eqnarray}
            L   & = & {189 \, q_0^3  \over 32 \pi^6 \, T^6}
               \int_0^{\pi} {\rm d} \theta \,
               \sin^3 \theta \int_0^{\infty}
               { {\rm d} q_x \over
               (q_x^2 + q_0^2)^{3/2}}
\nonumber      \\
               &  \times &
               \int_{q_r/2}^{\infty} {\rm d} \omega \,
               \left( \omega - {q_r \over 2} \right)^2
               \, { \omega^2 \over {\rm e}^{\omega / T}-1}.
\label{LRTfinal}
\end{eqnarray}

These equations take into account weak inelasticity of the $ef$
scattering, in analogy with
the neutrino bremsstrahlung due to the electron---nucleus
collisions (Haensel et al. 1996).

As
  can
be seen from Eq.\,(\ref{LRTfinal}),
$L$ depends on the
dimensionless parameter
 $t_0 = T/T_0$:
\begin{eqnarray}
      t_0
      &  \approx &
      0.00786 \, T_9
      \left( {m_p^\ast \, n_0
      \over m_p \; n_p} \right)^{1/2}
      \left[ 1 - \left( {T
      \over T_{cp}} \right)^4 \right]^{-1/2},
\label{t}
\end{eqnarray}
where $T_0 = T_p \sqrt{1 - (T/ T_{cp})^4}$, and
$T_p = \hbar \omega_p / k_{\rm B}$
is the  proton plasma temperature corresponding  to the
proton plasma frequency $\omega_p$ (Sect.~2).
Typically,
$T_p \sim 1$ MeV, under the conditions
  in
the NS cores.
  Let us analyze
two extreme cases:
$T \ll T_0$ and $T \gg T_0$.

\subsection{Low-temperature, bremsstrahlung regime, $T \ll T_0$}
In the limiting case of  $t_0 \ll 1$,  Eq.\,(\ref{LRTfinal}) gives
$L = \pi /4$. Then the neutrino emissivity is then given by
\begin{eqnarray}
    Q_{\rm flux}  &  =  & \frac{ \pi G_{\rm F}^2 e^2 \phi_0^2 C_+^2}
                          {9072 \, \hbar^9 c^8} \,
                          (k_{\rm B} T)^6 q_0
                          {\cal N}_{\rm F}
\nonumber       \\
                  &  \, &  \,
\nonumber       \\
               & \approx  & 2.09  \times 10^{16} \,
                  \bar{B}_{13} \, T_9^6 \,
                  \left({ n_p \, m_p
                  \over n_0 \, m_p^\ast }\right)^{1/2}
\nonumber       \\
                  & \times &
            \left[ 1 - \left( {T  \over T_{cp}}
            \right)^4 \right]^{1/2}
                    \; \;   {\rm erg~s^{-1}~cm^{-3}}.
\label{Qbrems}
\end{eqnarray}
The neutrino emission in the low--temperature regime
is very similar to that
due to the electron--nucleus bremsstrahlung
(e.g., Haensel et al. 1996). In particular,
  the
emissivity $Q_{\rm flux}$
is proportional to $T^6$. Therefore we will call this
regime the {\it bremsstrahlung regime}.

For the sake of completeness, we have also examined a
not
   too
realistic case of $q_0 \sim p_{\rm Fe}$
at $T \ll T_0$. The analysis is based on Eq.\, (\ref{Lgeneral}).
In this
   case,
the integration can be simplified,  and reduces to
\begin{equation}
          L   =   \int_0^1 \! {\rm d} y \frac{y_0^3}
               {(y^2 + y_0^2)^2} \;
               { 1 + y^2 \over 1 - y^2} \;
               \left[ 1 + \frac{2 y^2}{1 - y^2} \log (y) \right],
\label{Lfinal}
\end{equation}
where $y = q/(2 p_{\rm Fe})$. Now, $L$ is a function of
    the only
dimensionless parameter (see Sect.\~ 2)
\begin{eqnarray}
  y_0 & \equiv & { \hbar q_0 \over 2 p_{\rm Fe}}  =
       \left( { e^2 \over 3 \pi \hbar c}
       \, { p_{\rm Fe} \over m_p^\ast c} \right)^{1/2}
       \left[ 1 - \left( {T  \over T_{cp}} \right)^4 \right]^{1/2}
\nonumber \\
      & \approx & 0.0165 \left( {m_p \over m_p^\ast} \right)^{1/2}
       \left( {n_e \over n_0} \right)^{1/6}
       \left[ 1 - \left( {T  \over T_{cp}} \right)^4 \right]^{1/2}.
\label{ylambda}
\end{eqnarray}
In the limit of
$y_0 \ll 1$ we reproduce our basic result
$L = \pi/4$. For a finite $y_0$, the function $L$ given by
Eq.\,(\ref{Lfinal}) can be
   fitted
   (with
relative  error $< 2.4$ \%) by a simple expression
\begin{equation}
              L  = { \pi \over 4 \sqrt{1 + 0.7057 \, y_0^2}}.
\label{F1fit}
\end{equation}
%

%
\subsection{High-temperature, synchrotron
            regime, $T \gg T_0$}

In the case
   of
$t_0 \gg 1$,
   from Eq.\ (\ref{LRTfinal}) we obtain
an asymptotic form
$L \approx  (189 /\pi^6) \, \zeta(5)/t_0$,
where $\zeta(5) \approx 1.037$ is the  value of the Riemann zeta-function.
This yields
\begin{eqnarray}
    Q_{\rm flux}  &  =  & {\zeta(5) \,
              G_{\rm F}^2 e^2 \phi_0^2 C_+^2 \over
              12 \pi^6 \,  \hbar^8  c^7 } \, q_0^2 \,
              (k_{\rm B} T)^5 {\cal N}_{\rm F}
     \approx
 6.89  \times 10^{17} \, \bar{B}_{13}
\nonumber    \\
              & \times &
   T_9^5 \,
             { n_p \, m_p \over n_0  \,
             m_p^\ast }
             \left[ 1 - \left( {T  \over T_{cp}} \right)^4 \right]
             \; {\rm erg~s^{-1}~cm^{-3}}.
\label{highT}
\end{eqnarray}
The temperature dependence ($Q \propto T^5$) is the same as
for the synchrotron emission of neutrinos by
electrons (e.g., Kaminker et al. 1991).
Thus we will refer to  the high-temperature regime
as the {\it synchrotron
regime}.

Let us compare the  $ef$-scattering emissivity (\ref{highT})
with the emissivity $Q_{\rm syn}$
of the ``purely synchrotron process''
for the most important case of $\omega_B^\ast \ll T
\ll \mu^3 \omega_B^\ast$ (
$\mu_e$ is the electron chemical potential and $\omega_B^\ast =
eBc/\mu_e$ is the electron gyrofrequency). For
  the uniform magnetic field
  Kaminker et al. (1991) obtained:
\begin{eqnarray}
        Q_{\rm syn} & =  & {2 \zeta(5)   \over 9 \pi^5} \,
        {e^2  G_{\rm F}^2  C_+^2
        (k_{\rm B} T)^5 \, B^2 \over \hbar^8  c^7}
\nonumber    \\
        & \, & \,
\nonumber     \\
        &  \approx &  9.04 \times 10^{14} B^2_{13} T_9^5~~~
        {\rm ergs~cm^{-3}~s^{-1}}~.
\label{Qsynch}
\end{eqnarray}
%
   (The
numerical factor 9.04 differs from a factor
8.97 in Eq.\,(17)
of Kaminker et al. 1991, because in the present paper
 we use more accurate value of
   the Fermi
constant $G_{\rm F}$.)

     We
start with a brief discussion of the
$ef$ scattering at $t_0 \gg 1$
for  an arbitrary distribution of the fluxoid magnetic field $B(q)$.
In this case, we can set $q_r \to 0$ in
Eq.\,(\ref{LRT}). Then the integrations over $\theta$
and $\omega$ are decoupled from the integration over ${\bf q}$,
and
   can be
done analytically. This yields
the emissivity
\begin{equation}
               Q_{\rm flux} = \frac{1}{12 \pi^7 } \,
               \zeta(5) \, G_{\rm F}^2  C_+^2
               T^5 \, {\cal N}_{\rm F}
               \int {\rm d} {\bf q} \,
  B^2(q),
\label{QT5}
\end{equation}
where the integration
 should
be done over the entire $(q_x,q_y)$ plane.
The emissivity is seen to be independent
of the electron Fermi momentum.
In particular, in the limit of a (quasi) uniform magnetic field
$B(r)=B_0$, $B(q)= 4 \pi^2 \, B_0 \, \delta^{(2)} ({\bf q})$,
we can replace $B(q)^2 \to$
$4 \pi^2 \, B_0^2 \, \delta^{(2)} ({\bf q})
/ {\cal N}_{\rm F}$, and obtain
\begin{equation}
        Q_{\rm flux} = {\zeta(5)   \over 3 \pi^5} \,
         {e^2  G_{\rm F}^2  C_+^2
          (k_{\rm B} T)^5 \, B_0^2 \over  \hbar^8 c^7} .
\label{QT0}
\end{equation}

Comparing (\ref{QT0}) and (\ref{Qsynch}),
we
   have
$Q_{\rm syn} =$ \,\, $(2/3) \, Q_{\rm flux}$.
The same factor $2/3$  is obtained, if we treat $B$ in
$Q_{\rm syn}$ as a local magnetic field of a fluxoid,
average $Q_{\rm syn} \propto B^2$
over a lattice of fluxoids with the magnetic field
(\ref{Br}), and compare the result with $Q_{\rm flux}$,
Eq.\,(\ref{highT}).
Let us emphasize, that this emissivity $\bar{Q}_{\rm syn}$,
averaged over a nonuniform magnetic field, can differ from the
synchrotron emissivity $Q_{\rm syn}$ in the initial uniform
   field
$\bar{B}$ by a large factor of
$\overline{ B^2} / \bar{B}^2 = \phi_0 q_0^2 / (4 \pi \bar{B}) \sim$
$(d_{\rm F}/\lambda)^2$, where $d_{\rm F}$  is
an inter-fluxoid distance (Sect.~2).
This increase of $\bar{Q}_{\rm syn}$
is caused by the magnetic field enhancement within
the fluxoids due to magnetic flux conservation.
The enhancement is much stronger than the reduction of
neutrino-emission space  occupied by the fluxoids.

A detailed analysis shows, that
the difference by a factor of 2/3 comes from momentum space
available for  neutrino-pair momentum {\bf k}.
  The
space is different in the
case of synchrotron radiation in a {\it uniform magnetic field}
and in the case of $ef$ scattering
{\it by magnetic inhomogeneities}  (in our case
   ---
fluxoids).
Let us fix the electron pitch angle $\theta$
and average over the electron Larmor rotation.
In the both cases {\bf k} is concentrated to
the cone of a given $\theta$ (neutrinos are emitted
predominantly
  within a narrow angle interval $\delta \theta$
  in the direction of the electron pitch angle $\theta$),
and typically, $k \sim T$.
In the case of the $ef$ scattering,
  our estimates yield $\delta \theta_{\rm flux} \sim \sqrt{q_0/T}$.
The synchrotron radiation consists of many
  discrete cyclotron
harmonics $s$ with typical values $s \sim T / \omega_B^\ast \gg 1$,
for the quasiclassical regime of interest.
For a fixed $\theta$, the radiation
in each harmonics is peaked at the pitch-angle cone,
   and maximum
neutrino
   momenta
{\bf k}
   are restricted
   by a
parabolic surface (Kaminker et al.\ 1991). The maximum value of
$k$ in the pitch-angle direction is
$s \omega_B^\ast / (\sin \theta)^2$,
and,
   typically, $\delta \theta_{\rm syn}
   \sim s^{-1/3}$.
   Replacing a sum over harmonics by an integral we obtain a
   smoothed synchrotron {\bf k} space.
It fills only 2/3  of
the available momentum space,
due to the {\it parabolic momentum  space restriction}
for each harmonics
   (e.g., the surface area below the parabolic curve $y=1-x^2$
   at $0 \leq x \leq 1$ is exactly 2/3 of the surface
   area below a straight segment $y=1$).
A transition from the synchrotron
formula (\ref{Qsynch}) to the fluxoid formula (\ref{highT}) is expected
to occur at
   $\delta \theta_{\rm flux} \sim \delta \theta_{\rm syn}$.
This is equivalent to the condition
$\lambda \sim (c/ \omega_B^\ast) s^{-1/3}$, which takes place when
the scale of the magnetic field variation
   $\lambda$
becomes comparable to a distance
   along
which a neutrino
pair is emitted in a uniform magnetic field, i.e.,
to the $s^{1/3}$-th part of
the Larmor radius $c/\omega_B^\ast$.

%
\subsection{Overall fit and overview}
Combining the results of Sects.~4.1 and 4.2, we can propose
the following fitting  expression for the quantity
$L$ in Eq.\,(\ref{QthrL})
   at $y_0 \ll 1$:
\begin{equation}
    L = L_0 \, U \, V,
\label{F1F2}
\end{equation}
\begin{eqnarray}
             L_0 & = & {\pi \over 4} \,
             {0.260 \, t_0 + 0.0133  \over
             t_0^2 + 0.25 \, t_0 + 0.0133},
  ~~~~U
  =
  {2 \gamma+1  \over 3 \gamma+1 },
\nonumber \\
  \gamma & = & \left({ \hbar \omega_B^\ast  \over \mu_e} \right)^2
             { t_0 \over y_0^2}
             \approx \, 8.38 \times 10^{-12} \, T_9
             \bar{B}_{13}^2
\nonumber   \\
             &  \times &
             \left( { n_0 \over n_e } \right)^{13/6} \,
             \left( {m_p^\ast \over m_p } \right)^{3/2} \,
             \left[ 1 - \left( {T \over T_{cp}} \right)^4 \right]^{-3/2}.
\label{2/3} \\
         V & = & 1 + {4 \pi \bar{B} \lambda^2 \over \phi_0}
\nonumber     \\
        & \approx & 1 + 0.00210 \, \bar{B}_{13} \,
       {m_p^\ast \, n_0 \over m_p~\, n_p}
       \left[ 1 - \left( {T \over T_{cp}} \right)^4 \right]^{-1}.
\label{unif}
\end{eqnarray}
Here, $L_0$ is the analytic expression
which fits the results of our numerical calculations
of $L$ from Eq.\,(\ref{LRTfinal})
with error $<$ 1 \% for any $t_0$.
The factor $U$ ensures, somewhat arbitrarily,
the difference by a factor of 2/3 between the cases of
weakly and strongly nonuniform magnetic
  fields
(Sect.~4.2).
The factor $V$ provides a
smooth transition from the $ef$ scattering to the pure
synchrotron emission in the case when a fluxoid radius
$\lambda$ becomes
very large ($T \to T_{cp}$), the neighboring fluxoids overlap and the
overall magnetic field becomes nearly uniform.

Using our results, we can  follow the evolution of $Q_{\rm flux}$
in the course of the superconductivity onset. If $T \geq T_{cp}$,
the emissivity $Q_{\rm flux}$
   is given by Eq.\ (\ref{Qsynch}) since it
is essentially the same
as the synchrotron emissivity
$Q^{(0)}_{\rm syn}=Q_{\rm syn}(\bar{B})$ in a locally uniform
`primordial' magnetic field $B=\bar{B}$.
After $T$ falls only slightly below $T_{cp}$,
the factor $U$ transforms from $U=2/3$ (`pure synchrotron')
to $U=1$ ($ef$ scattering in the synchrotron regime).
The fluxoid structure is still not very pronounced, i.e.,
$V \approx 4 \pi \bar{B} \lambda^2 / \phi_0 $,
$Q_{\rm flux} \sim Q_{\rm syn}^{(0)}$.
When $T$ decreases to about $0.8 \, T_{cp}$,
the fluxoid radius $\lambda$ nearly achieves its zero-temperature
value $\lambda_0$, and
$V \simeq 1$. Accordingly, $Q_{\rm flux} /Q^{(0)}_{\rm syn}$ grows
by a factor of
$(d_{\rm F}/\lambda_0)^2$ due to the magnetic field
confinement within the fluxoids (Sect.~4.2). Simultaneously,
$t_0$ decreases from very large values at  $T
\to T_{cp}$
to  $t_0\simeq T_{cp}/T_p$.

Let us first consider  the case in which the proton
plasma temperature is $T_p \ll T_{cp}$.
Then the emissivity
$Q_{\rm flux}$ is enhanced  with respect to $Q^{(0)}_{\rm syn}$
by a factor of $(d_{\rm F}/\lambda_0)^2$ at $T \approx 0.8 T_{cp}$,
and this enhancement remains nearly constant over a wide temperature
range down to $T_p$. Within this range, the $ef$ scattering
operates in the synchrotron regime.
At lower temperatures, $T \la T_p$,
the synchrotron regime transforms into the bremsstrahlung regime
(Sect.\ 4.1) and we have $Q_{\rm flux} \sim Q^{(0)}_{\rm syn}
(d_{\rm F}/\lambda_0)^2 \, (T/T_p)$.
Thus, for $T\la T_p$ the emissivity $Q_{\rm flux}$
decreases with respect to $Q^{(0)}_{\rm syn}$, and
may become lower
than $Q^{(0)}_{\rm syn}$.

In the opposite case of $T_p \ga T_{cp}$, the $ef$ scattering operates
in the synchrotron regime only in a narrow
temperature range below $T_{cp}$, and transforms  into
the bremsstrahlung regime at lower $T$. In the latter regime, we
have
$Q_{\rm flux} \sim $
$Q^{(0)}_{\rm syn}(d_{\rm F}/
    \lambda
)^2 (T/T_p)$.

%
\section{Neutrino bremsstrahlung due to $ep$ and $ee$ scattering}
As
   will
be shown in Sect.~6,
the $ef$ scattering  mechanism can be important,
if protons and neutrons in a NS core
are highly superfluid (have
   high
critical
temperatures $T_{cp}$, $T_{cn}$),
    so that
the rates of the ``standard''
Urca and nucleon bremsstrahlung processes are strongly suppressed.
    Under
such conditions,
 one should carefully take into account {\it all} the neutrino
processes, even if they are negligible
 for relatively low critical temperatures.

In this section, we consider two additional neutrino
production mechanisms, the neutrino-pair
bremsstrahlung due to the $ep$ and $ee$ scattering:
\begin{eqnarray}
    e + p \to e + p + \nu + \bar{\nu},
\nonumber \\
    e + e \to e + e + \nu + \bar{\nu}.
\label{newnu}
\end{eqnarray}
As far as we know,
these mechanisms were neglected in the
neutron-star cooling
    simulations.

    Let us
restrict ourselves to simple estimates
of the emissivities $Q_{ep}$ and $Q_{ee}$; detailed
    analysis will be published elsewhere
(Kaminker et al.\ in preparation).
    We
start from the well-known
emissivity $Q_{eZ}$ of the electron-nucleus
bremsstrahlung (e.g., Haensel et al. 1996).
   In
the particular case of the $ep$ collisions
of relativistic, strongly degenerate electrons in
a nonsuperfluid matter, we
     obtain
\begin{equation}
     Q^{(0)}_{ep} = { 8 \pi C_+^2 e^4 G_{\rm F}^2 \over 567 \, \hbar^9 c^8}
     \, (k_{\rm B}T)^6 n_p \,L_{ep}~,
\label{Q0ep}
\end{equation}
where $L_{ep}$ is the Coulomb logarithm.
  The above formula has been derived assuming,
that an energy transfer to a proton in a neutrino-emission act is
much smaller than $T$. As applied to our case
of strongly degenerate
   protons, the latter means
that the Pauli principle imposes
no restrictions on the states of
   protons. This
is true
as long as $T \gg q_{\rm s}$, where $q_{\rm s}$ is
an inverse length of
   plasma screening
of the Coulomb $ep$
interaction
(see below). However, in reality we have
just an opposite regime, $T \ll q_{\rm s}$, and  Eq.
(\ref{Q0ep}) does not hold.

 To estimate $Q_{ep}$ at $T \ll q_{\rm s}$,  we assume
that $Q_{ep} \propto \nu_{ep}$, where $\nu_{ep}$
is some effective $ep$ collision frequency.
If our assumption is correct, we have
$Q_{ep} \approx Q^{(0)}_{ep} \, \nu_{ep} / \nu^{(0)}_{ep}$,
where $\nu^{(0)}_{ep}$ and $\nu_{ep}$  refer to the high- and
low- temperature cases, respectively.
For high temperatures, we will use the effective
frequency of nearly elastic collisions $\nu^{(0)}_{ep}$
(e.g., Yakovlev and Urpin 1980)
which determines the electric and thermal
conductivities of degenerate electrons. For $T \ll q_{\rm s}$,
we employ the collision frequency $\nu_{ep}$ that
determines the electron thermal conductivity
(e.g., Gnedin and Yakovlev 1995). Performing the above
rescaling, we obtain
\begin{eqnarray}
     Q_{ep} & = & { \pi^5 G_{\rm F}^2 e^4 C_+^2  m_p^{\ast2} \over
           945 \hbar^9 c^8 y_{\rm s}^3 p_{\rm Fe}^4 } \, n_p \,
           (k_{\rm B} T)^8 R_{ep}
\nonumber \\
    & \approx &  {3.42 \times 10^{14} \over y_{\rm s}^3}
          \left({ m_p^\ast \over m_p } \right)^2
          \left( { n_0 \over n_p} \right)^{1/3}
\nonumber    \\
            & \times &
          R_{ep} \,  T_9^8~~{\rm ergs~cm^{-3}~s^{-1}},
\label{Q_ep}
\end{eqnarray}
where $y_{\rm s} = \hbar q_{\rm s} /(2 p_{\rm Fe})$
is the plasma screening
parameter.
We additionally introduce the factor
$R_{ep}$ that describes suppression of the emissivity
by the proton superfluidity. According to Gnedin \& Yakovlev
(1995) the plasma screening in $npe$ matter of the NS
cores is defined by
\begin{equation}
    y_{\rm s}^2 = { e^2 \over \pi \hbar c }
    \left(1 + { m_p^\ast \, p_{\rm Fp} \over m_e^\ast \, p_{\rm Fe}}
    Z_p \right),
\label{y_s}
\end{equation}
where $m_e^\ast = \mu_e / c^2$.
The first term in brackets comes from
the electron screening while the second is due to the proton
screening. The latter contains the factor $Z_p$ which
describes reduction of the proton screening by the
proton superfluidity. According to Gnedin \& Yakovlev (1995)
this factor is fitted as
\begin{eqnarray}
   Z_p & = & \left(0.9443 + \sqrt{(0.0557)^2 + (0.1886 v)^2}
              \right)^{1/2}
\nonumber    \\
             &  \times  &
             \exp \left(1.753 - \sqrt{(1.753)^2+ v^2}\right),
\label{Z_p}
\end{eqnarray}
where $v$ is the proton superfluid gap parameter
\begin{eqnarray}
   v & = & { \Delta_p(T) \over k_{\rm B} T } =
   \sqrt{1 - { T \over T_{cp} } }
\nonumber    \\
         & \times   &
        \left(1.456 - 0.157 \, \sqrt{T_{cp} \over T}  +
         1.764 \, {T_{cp} \over T}  \right).
\label{gap}
\end{eqnarray}
If protons are normal ($T \geq T_{cp}$), one has
$R_{ep}=Z_p=1$ in Eqs.\ (\ref{Q_ep}) and (\ref{y_s}).
The factor $R_{ep}$, which damps  $Q_{ep}$ at $T < T_{cp}$,
should be the same as the factor $R_{np}$
that describes reduction of the neutrino emission in $np$ collisions
by the proton superfluidity (Yakovlev and Levenfish 1995):
\begin{eqnarray}
     R_{ep} &  =  & {1 \over 2.732} \left[
       a \exp \left(1.306 - \sqrt{(1.306)^2+v^2 }\right) \right.
\nonumber \\
       & + & \left.
     1.732 \,
       b^7 \exp \left(3.303 - \sqrt{(3.303)^2+4 v^2 }\right) \right],
\label{R_ep}
\end{eqnarray}
with $a=0.9982+ \sqrt{(0.0018)^2+(0.3815v)^2}$,  \\
$b=0.3949+ \sqrt{(0.6051)^2+(0.2666v)^2}$.

Now consider the neutrino emissivity
$Q_{ee}$ due to $ee$ scattering. Under the same assumption,
we obtain $Q_{ee} = Q_{ep} \, \nu_{ee} / \nu_{ep}$,
where $\nu_{ee}$ is the effective $ee$ collision frequency
that
   determines
the electron thermal
conductivity (e.g., Gnedin \& Yakovlev 1995). Then
\begin{eqnarray}
     Q_{ee} & = & { \pi^5 G_{\rm F}^2 e^4 C_+^2  \over
           378 \hbar^9 c^{10} y_{\rm s}^3 p_{\rm Fe}^2 } \, n_e \,
           (k_{\rm B} T)^8
\nonumber \\
    & \approx &  {1.07 \times 10^{14} \over y_{\rm s}^3}
          \left( { n_e \over n_0} \right)^{1/3} \,
          T_9^8~~{\rm ergs~cm^{-3}~s^{-1}}.
\label{Q_ee}
\end{eqnarray}

Although we do not calculate $Q_{pe}$ and $Q_{ee}$
exactly, we believe that Eqs.\ (\ref{Q_ep}) and (\ref{Q_ee})
give reliable estimates. The emissivity $Q_{ep}$
is reduced exponentially by the strong proton
superfluidity, while $Q_{ee}$ is affected by the superfluidity
  in a
much weaker manner, only through the plasma screening
parameter (\ref{y_s}).
If protons are normal they provide the major contribution
into the plasma screening. If they are strongly superfluid,
a weaker electron screening becomes important,
which {\it enhances} $Q_{ee}$, but not
to a great extent (see below).

%
\section{Discussion}
Figures 1 and 2 illustrate the efficiency of various
neutrino production mechanisms in $npe$ matter of
the NS cores. We compare the neutrino
emissivity due to the $ef$
scattering or due to the electron synchrotron radiation
(Sect.~4) with those produced by the neutron
and proton branches of the modified Urca reactions
(Friman \& Maxwell 1979, Yakovlev \& Levenfish 1995),
\begin{eqnarray}
  n + n \to n + p + e + \bar{\nu}_e,~~
  n + p + e \to n + n + \nu_e~
  ;
\nonumber \\
  n + p \to p + p + e + \bar{\nu}_e,~~
  p + p + e \to n + p + \nu_e~;
\label{Murca}
\end{eqnarray}
by the nucleon-nucleon bremsstrahlung processes
(Friman \& Maxwell 1979, Yakovlev \& Levenfish 1995)
\begin{eqnarray}
    n + n \to n + n + \nu + \bar{\nu},
\nonumber \\
    n + p \to n + p + \nu + \bar{\nu},
\nonumber \\
    p + p \to p + p + \nu + \bar{\nu};
\label{NN}
\end{eqnarray}
and also by the $ep$ and $ee$ bremsstrahlung processes
(\ref{newnu}) considered in Sect.~5.
For better presentation, the total emissivity
of the both branches of the modified Urca reactions
(\ref{Murca}) is represented
by  a single curve {\it `Murca'}, and the total emissivity
 from the three nucleon-nucleon bremsstrahlung
reactions (\ref{NN}) are  represented
by  a single  curve {\it `NN'}.
We adopted a moderately
stiff equation of state of Prakash et al. (1988)
(the same version as used by Page \& Applegate 1992)
and choose matter density $\rho = 5.6 \times 10^{14}$ g cm$^{-3}$
($n_p = n_e \approx 0.0207$ fm$^{-3}$, $n_n \approx 0.323$
fm$^{-3}$),
well below the threshold density $1.30 \times 10^{15}$ g cm$^{-3}$
at which the direct Urca process becomes allowed (Lattimer et al.
1991). Therefore our curves correspond to the {\it standard}
neutrino emission (no exotic cooling agents, see, e.g.,
Pethick 1992). We show temperature dependence of
various neutrino  emissivities in the range from
$10^8$ K to $ 5 \times 10^9$ K, which
    is
the most important for the cooling theories of NS.
The nucleon effective masses are set equal to 0.7
of the masses of bare particles.
The neutrino emissivities vary slowly with density
(below the direct Urca threshold), i.e., we present a typical
situation within the NS cores.
In both figures, we assume the presence
of the `primordial' magnetic field $\bar{B}= 10^{12}$, $10^{13}$
or $10^{14}$ G.
\begin{figure}
\begin{center}
\leavevmode
\epsfxsize=8.7cm \epsfbox{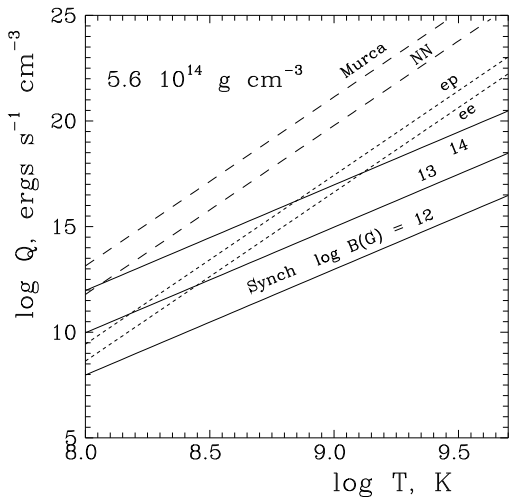}
\end{center}
\caption[ ]{
Neutrino energy loss rates from various processes,
versus temperature in a non--superfluid $npe$ matter
of density $5.6 \times 10^{14}$ g cm$^{-3}$.
{\it Murca} --- total contribution of two branches of the
modified Urca process, Eq.~(\ref{Murca});
{\it NN} --- total contribution
 of three nucleon bremsstrahlung processes,
 Eq.~(\ref{NN}); $ep$ and $ee$ ---  bremsstrahlung processes,
 Eq.~(\ref{newnu}),
due to the $ep$ and $ee$ scattering, respectively;
solid lines --- synchrotron radiation of electrons for
three values of the magnetic field.
}
\label{fig1}
\end{figure}

Figure 1 corresponds to a nonsuperfluid matter.
 If the magnetic field
is absent, the Murca processes are seen to be the most important
ones,
in the displayed temperature range. The $NN$ bremsstrahlung
is much weaker, and the $ep$ and $ee$ bremsstrahlung processes
are even weaker.
The emissivities
  of
all these mechanisms vary as $T^8$,
whereas the emissivity of the
synchrotron radiation
varies as $T^5$. Thus the synchrotron radiation
is negligible for $T \ga 10^9$ K but becomes
more important with decreasing $T$. At $T \sim 10^8$ K
and $B=10^{14}$ G the
   synchrotron
dominates over all other mechanisms except the Murca
processes, and
at slightly lower $T$ it becomes the most efficient
of all mechanisms. However, such
temperatures are too low to be important for the
cooling theories: the neutrino luminosity of NS
becomes then smaller than the photon luminosity
from the stellar surface, and the neutrino emission looses
its
   significance
as a source of the NS cooling.
\begin{figure}
\begin{center}
\leavevmode
\epsfxsize=8.7cm \epsfbox{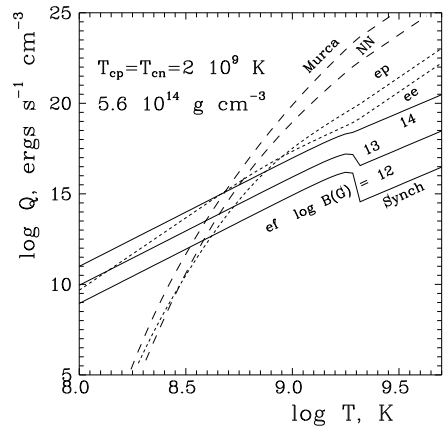}
\end{center}
\caption[ ]{
Same as in Fig.~1 but in the presence of
proton and neutron superfluidities with
$T_{cp}=T_{cn}= 2 \times 10^9$ K. At $T<T_{cp}$
the initially uniform magnetic field splits into
fluxoids, and the synchrotron emission transforms
into the emission due to the $ef$
scattering. The nucleon superfluidity suppresses
all mechanisms but the $ee$ and $ef$
scattering.
}
\label{fig2}
\end{figure}
%
Figure 2 shows how the neutrino emission
displayed in Figure 1 is modified
by the nucleon superfluidity.
In our example, the protons and neutrons
become superfluid at $T_{cp}=T_{cn}= 2 \times 10^9$ K.
These are typical, moderate critical temperatures
for the nucleon superfluidity (e.g., Takatsuka \& Tamagaki 1993).
The factors which describe superfluid suppression
of the modified Urca processes (\ref{Murca}) and $NN$ collisions
(\ref{NN}) are taken from Yakovlev \& Levenfish
(1995). At $T \la 10^9$ K the nucleon
superfluidity is seen to
  strongly
reduce all the
neutrino production processes which involve nucleons.
If the magnetic field were absent, the $ee$ scattering
would be the main neutrino generation mechanism at
$T \la 5 \times 10^8$ K. However,  the $ef$
scattering in the presence of $\bar{B} \ga
10^{13}$ G also becomes important, and it can even dominate.
After the superfluidity onset ($T < T_{cp}$),
the magnetic field splits into fluxoids
and the synchrotron radiation transforms into the
radiation due to the $ef$ scattering.
This enhances $Q_{\rm flux}$
  owing
to the field amplification
within the fluxoids (Sect.~4). The enhancement
is more pronounced at lower $\bar{B} \sim 10^{12}$ G
at which the field amplification is stronger. This is
a rare case in which the neutrino emissivity {\it increases} with
decreasing $T$. For the conditions displayed in Fig.~2,
the proton plasma temperature $T_p =5.47 \times 10^{10}$ K
is much higher than the superfluid critical temperature $T_{cp}$.
Therefore, the synchrotron regime (Sect.~4.2) in the $ef$
scattering operates only at $T \la T_{cp}$, during the phase
of the fluxoid formation.
Very soon after $T$ falls below $T_{cp}$, the synchrotron
regime transforms into the bremsstrahlung regime (Sect.~4.1)
which operates further with decreasing $T$.

We have analysed a number of cases,
varying the proton and neutron
critical temperatures $T_{cp}$ and $T_{cn}$.
Our principal conclusion is that the standard
neutrino production mechanisms (\ref{Murca}) and
(\ref{NN}) dominate in the temperature domain
of practical interest if either $T_{cp}$ and/or $T_{cn}$
are not too high (not higher than about
$10^9$ K). If, however, both critical temperatures
are high, the situation is similar to that
shown in Fig.~2.
The superfluidity suppresses all the
traditional neutrino generation mechanisms, and the main
neutrino production at $T \la 5 \times 10^8$ K comes
either from the $ef$ or the $ee$ scattering.

In principle, neutrinos can be generated
also in the $pp$ collisions
of {\it normal} protons in the non-superfluid cores of the
fluxoids (Sect.~2) as well as
in  the $en$ collisions. However, the non-superfluid cores
are very thin, the
emission volume is minor, and  the first process is
inefficient. The neutrino bremsstrahlung due
to $en$ scattering is
  also inefficient
since it occurs through electromagnetic interaction
involving neutron magnetic
moment (e.g., Baym et al. 1969) and since it is suppressed
by the neutron superfluidity.

While it is commonly accepted that protons form a type II 
superconductor, a possibility that they actually form a type 
I superconductor cannot be excluded. This uncertainty results 
from the  lack of  a precise knowledge of the nucleon-nucleon 
interaction in dense nuclear matter, and from the approximations 
and deficiencies  of the many-body theory of dense nucleon 
matter. In the case of a type I proton superconductor, which 
  corresponds  to  $\xi_0>\sqrt{2}\lambda_0$ (de Gennes 1966), 
 cooling below  $T_{cp}$ is expected to be accompanied by a 
transition of the magnetized interior to  an ``intermediate 
state'' (de Gennes 1966, Baym et al. 1969). The ``intermediate 
state'' would consist of alternating regions of normal matter 
containing  magnetic flux, and superconducting regions exhibiting 
a complete expulsion of magnetic flux. The specific spatial structure 
of the ``intermediate state'' would result from the condition of 
the minimum of the thermodynamic  potential at a fixed macroscopic 
magnetic flux (de Gennes 1966). Qualitatively, we expect that 
 the presence of an ``intermediate state'' in the superconducting 
proton core will imply an enhancement of its electron synchrotron 
$\bar\nu\nu$ emissivity, compared to the case of a normal proton 
core,  because  of $\overline{B^2}>(\bar{B})^2$. However, for 
$\bar{B}\la 10^{14}~{\rm G}$ this enhancement will be much 
smaller  than that characteristic of the type II superconductor, 
 in which magnetic field is confined to fluxoids. 
%
\section{Conclusion}
We have considered (Sects.~3 and 4) the neutrino-pair radiation
(\ref{ef}) due to
scattering of relativistic, degenerate electrons off
threads of quantum magnetic flux --- fluxoids --- in
 the $npe$ matter within  the
superfluid NS cores.
We have shown that this mechanism is a natural
 generalization of the synchrotron emission of neutrinos by
electrons in a non-superconducting matter
(with a locally uniform magnetic field) to the case in which the
protons become superfluid and the magnetic field
splits into the fluxoids. According to our results, this
mechanism can operate either in the synchrotron
(Sect.~4.2) or in the bremsstrahlung (Sect.~4.1) regime.
We have obtained a simple fitting  expression
(\ref{F1F2}) which
  reproduces
the emissivity (\ref{QthrL}) for any
parameters of practical interest.

Furthermore, we have estimated (Sect.~5) the neutrino emissivities
in two additional neutrino production mechanisms (\ref{newnu}),
the $ep$ and $ee$ bremsstrahlung processes.
In a non--superfluid matter, these mechanisms
are much weaker than
the standard neutrino emission mechanisms,
such as the modified Urca processes (\ref{Murca}) and
the nucleon--nucleon bremsstrahlung (\ref{NN}).
If, however, both
superfluid critical temperatures, $T_{cp}$ and $T_{cn}$,
are higher than  10$^9$ K, then
the superfluidity  strongly suppresses  all the traditional
(standard)
neutrino energy losses. In such a case, the
main neutrino generation in a NS core at
$T \la 5 \times 10^8$ K occurs either via the
electron-fluxoid or via the
$ee$ scattering (Sect.~6).
Therefore our results can be of
particular importance for simulating the cooling of
 NSs
with
highly superfluid cores.

\acknowledgements
The authors are grateful to Kseniya Levenfish who
participated at the initial stage of this work.
One of the authors (ADK) acknowledges excellent working
conditions and hospitality of N. Copernicus Astronomical
Center in Warsaw.
This work was supported in part by the
RBRF (grant No. 96-02-16870a), INTAS (grant No. 94-3834),
DFG-RBRF (grant No. 96-02-00177G), and the KBN grant
 2P 304 014 07.

\end{document}